\documentclass[a4paper]{article}

\usepackage{amsmath,amssymb,amstext,graphicx}

\newcommand{\ads}{\ensuremath{\mathrm{AdS}}}
\renewcommand{\S}{\mathrm S}
\newcommand{\SU}{\mathrm{SU}}
\newcommand{\U}{\mathrm{U}}
\newcommand{\N}{\mathcal N}

\newcommand{\dd}{\mathrm d}
\newcommand{\R}{\mathfrak R}
\newcommand{\T}{\mathcal T}
\renewcommand{\Im}{\mathop\mathrm{Im}}
\renewcommand{\L}{\mathcal{L}}
\newcommand{\del}{\partial}

\title{\vspace{-1.5em}\parbox{\textwidth}{\small\hfill MPP--2009--9}\\[1.5em]
QGP thermodynamics and meson spectroscopy with AdS/CFT}
\author{\parbox{\textwidth}{\centering \mbox{Johanna Erdmenger${}^a$\footnote{jke@mppmu.mpg.de}\,, Matthias Kaminski${}^b$\footnote{matthias.kaminski@uam.es}\,, Felix Rust${}^a$\footnote{rust@mppmu.mpg.de}}}\\[\bigskipamount]
\small
\parbox{\linewidth}{%
\centering
${}^a$\;Max-Planck-Institut f\"ur Physik (Werner-Heisenberg-Institut),\\ F\"ohringer Ring 6, 80805 M\"unchen, Germany\\[\smallskipamount]
${}^b$\;Instituto de Fisica Te\'orica, UAM/CSIC, Facultad de Ciencias,\\ C-XVI Universidad Aut\'onoma de Madrid Cantoblanco,\\ Madrid 28049, Spain
}
}
\date{}

\begin{document}
\maketitle

\begin{abstract}%

	\noindent In this talk we present applications of the \ads/CFT
	correspondence to strongly coupled systems at finite temperature and
	particle density. The model we investigate contains adjoint matter described
	by the gauge multiplet of $\N=4$, as well as fundamental matter given by the
	hypermultiplet of $\N=2$ Super Yang-Mills theory. In some aspects these
	systems can be thought of as models for the quark-gluon plasma.
	
	In the first part we review some properties of meson spectra obtained from
	these holographic models. We discuss the implications of finite temperature
	and particle density in these string-theory motivated setups. In particular,
	we find a broadening of the vector meson peaks in the relevant spectral
	function at finite density. However, we do not observe a movement of the
	resonances to lower frequencies.
	
	Moreover, we analyze the effects of strong coupling on heavy meson diffusion
	in medium. To do so we make use of an effective model with dipole
	interaction, which is valid for heavy quarks at arbitrary coupling
	strength. We calculate the momentum broadening\,---\,normalized to the
	in-medium mass shift\,---\,and compare the large 't~Hooft coupling \ads/CFT
	result with a perturbative result for weak coupling. We find that the
	momentum broadening is reduced at large 't~Hooft coupling, leading to
	increased relaxation time.
	
	\vfill
	
	\noindent
	\footnotesize
	Based on a talk given by Felix Rust at the \emph{8th Conference ``Quark
	Confinement and the Hadron Spectrum'', September 1st-6th 2008, Mainz, Germany.}
	
\end{abstract}

\newpage 


\section{Introduction}
	
	The \ads/CFT correspondence maps string theory, or certain limits thereof,
	to quantum field theory. A well-established explicit formulation maps
	classical supergravity to large $N$ gauge theory. Application of the
	\ads/CFT correspondence to strongly coupled systems at finite temperature
	and particle density is interesting from two points of view. First,
	holographic models allow for qualitative comparison of string-theory
	motivated calculations with phenomenology. Second, the \ads/CFT
	correspondence provides a very useful approach to dynamical processes
	in strongly coupled field theories. Holographic models therefore may prove
	to be valuable tools to gain insight into the nature of matter under extreme
	conditions, as it existed shortly after the big bang or in heavy ion
	collisions.
	
	We pursue the strategy of the top--down approach. This means we use
	solutions of ten-dimensional supergravity to investigate the properties of
	the dual field theories. In this way we restrict our attention to the best
	accessible field theory limits, i.\,e.\ asymptotic low energy, large number
	$N$ of color degrees of freedom and strong coupling.
	
	In particular we are interested in bound states of fundamental and
	anti-fundamental matter described by the hypermultiplet of $\N=2$ Super
	Yang-Mills (SYM) theory, accompanied by gauge fields in the adjoint gauge
	multiplet of $\N=4$ SYM theory. We interpret these bound states as
	quark-antiquark mesons in a quark-gluon plasma. In the quantum field theory,
	the vector valued bound states of quarks can be described by the retarded
	two-point correlation function $G^R$ of flavor currents $J$,
	\begin{equation}
	\label{eq:retardedGreen}
		G^R(\omega) = -i\int\!\dd^4x\; e^{i\,\vec k\vec x}\,\theta(x^0)\,\big<[J(\vec 0),J(\vec x)]\big>.
	\end{equation}
	Generalizations of the \ads/CFT correspondence allow to compute these
	correlators at \emph{strong coupling}, and to derive meson mass spectra from
	them.
	
	The same methods can be used to estimate the effect of strong coupling on
	meson diffusion. The coupling constant in our effective model is determined
	by the quotient $\lambda/m_q$ of 't~Hooft coupling and quark mass. We
	compare momentum broadening at \emph{strong coupling}, obtained from
	\ads/CFT calculations, to perturbative results at \emph{weak coupling}.
	These efforts are motivated by the search for a mechanism of the suppression
	of heavy quarks observed at RHIC, and energy loss of mesons in the strongly
	coupled medium.

\section{Holographic meson spectra at finite temperature and particle density}
	
	We work in the D3/D7 setup, where a finite small number of $N_f$ D7-branes is
	embedded into the $\ads_5\times\S^5$ black hole background, which is
	generated by a stack of $N\gg N_f$ D3-branes. We restrict to the low energy
	limit, where the resulting degrees of freedom of string excitations are dual
	to the $\N=4$ gauge multiplet plus the $\N=2$ hypermultiplet of SYM theory.
	We mainly focus on the latter degrees of freedom which transform in the
	fundamental representation of $\SU(N)$ and $\SU(N_f)$. We may therefore
	interpret $N$ as the number of colors, $N_f$ as the number of flavors, and
	the hypermultiplet as the supersymmetric realization of quark fields.
	
	According to the \ads/CFT dictionary, the parameters and fields of the
	gravity setup translate into field theory equivalents as follows. The \ads\
	black hole radius is proportional to the field theory temperature $T$, which
	breaks supersymmetry. The separation of the D3- and D7-branes is
	proportional to the quark mass $m_q$, which breaks conformal symmetry. We
	normalize dimensionful quantities to the temperature and use $m\propto
	m_q/T$ as a free parameter. The second free parameter is the
	particle density $\tilde d$, which is proportional to the value of the time
	component of the $\SU(N_f)$ valued gauge field $A$ in the supergravity
	theory, evaluated on the boundary of $\ads_5$ space. Additional to the
	flavor symmetry group $\SU(N_f)$ the full symmetry group
	$\U(N_f)\cong\U(1)\times\SU(N)$ of the brane configuration also exhibits a
	$U(1)$ factor. We work in $N_f=2$ and interpret the $\U(1)$ charge as baryon
	number and the $\SU(2)$ charge as isospin number. In this way we can tune
	the baryon and isospin particle density $\tilde d$ by specifying the
	boundary value of $A$ \cite{Kobayashi:2006sb,Erdmenger:2007ja}.
	
	To derive the spectra of bound states of quark-antiquark pairs as above, we
	need to determine the embedding of the D7-branes as well as the gauge field
	configuration on the brane. They are determined by extremizing the
	Dirac-Born-Infeld action, which is given by
	\begin{equation}
	\label{eq:DBIaction}
		S_\text{DBI} = -T_7\int\dd^8\xi\;\sqrt{G+2\pi\alpha' F}.
	\end{equation}
	Here the D7-brane embeddings determine the induced metric $G$ of the
	D7-brane, and $F=\dd A$ is the field strength tensor of the gauge field on
	the brane, and $T_7$ is the brane tension.
	
	We describe the field theory meson spectrum in terms of the spectral
	function $\R$, defined as
	\begin{equation}
		\R(\omega) = -2 \Im G^R(\omega).
	\end{equation}
	To obtain $G^R$, as defined in \eqref{eq:retardedGreen}, we make use of the
	\ads/CFT correspondence which allows us to identify the partition sum of the
	D3/D7 setup with the generating functional of the dual field theory. The
	partition function is given by $Z=\exp S_{\text{DBI}}$. As the boundary
	value of the gauge field $A$ is the source of the operator $J$ in
	\eqref{eq:retardedGreen}, we can obtain correlation functions by evaluating
	functional derivatives of the above partition function \cite{Son:2002sd}.
	Explicit calculation \cite{Myers:2007we} yields
	\begin{equation}
		G^R = \frac{N_f N T^2}{8}\,\lim_{\rho\to\infty}\left(\rho^3\frac{\del_\rho A}{A}\right),
	\end{equation}
	where we have to insert $A$, which we obtain as a numerical solution to the
	equations of motion derived from the DBI action \eqref{eq:DBIaction}
	\cite{Erdmenger:2007ja}.
	
\begin{figure}
		\newlength{\figWidth}
		\setlength{\figWidth}{.3\textwidth}
		\newlength{\picHeight}
		\setlength{\picHeight}{0mm}
		\begin{minipage}[t]{\figWidth}
			\rule{0mm}{\picHeight}%
			\includegraphics[width=\linewidth]{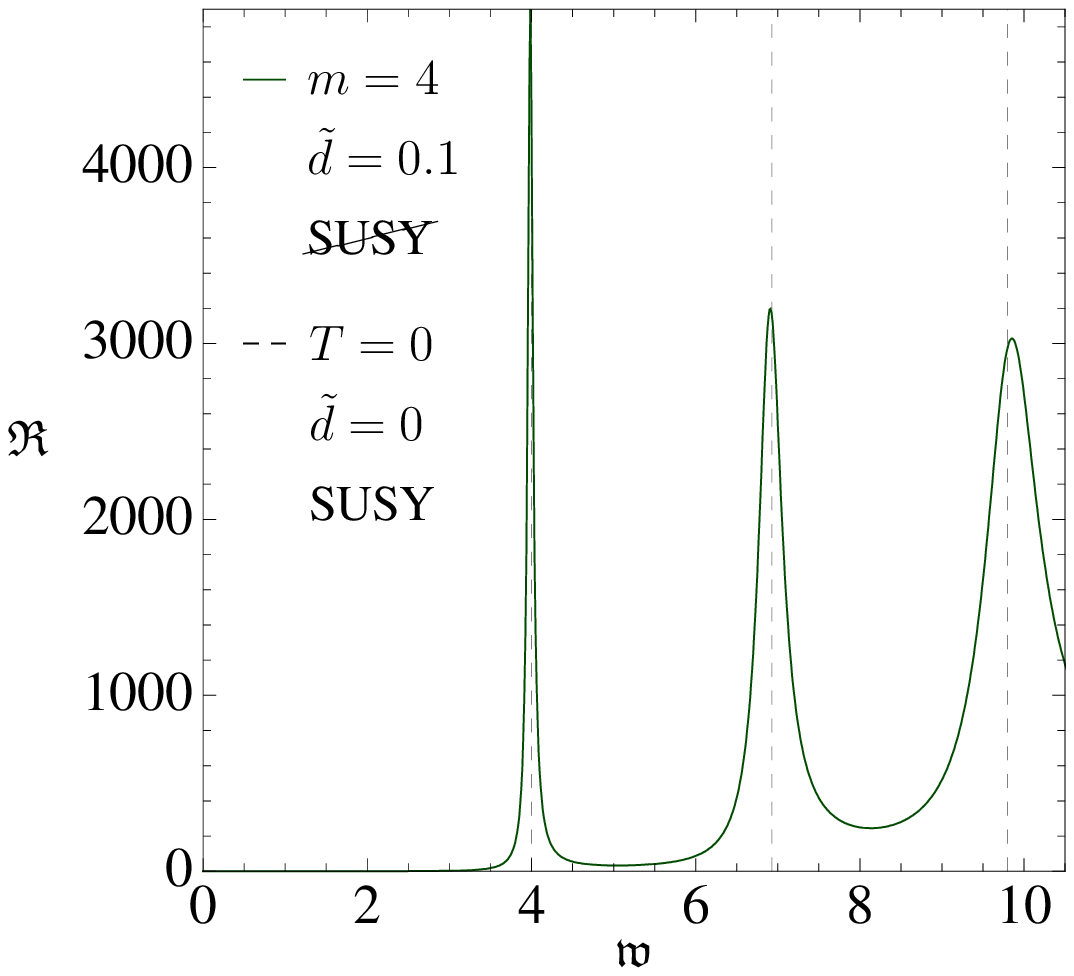}
			\caption{At low temperatures the spectrum coincides with the exact
				supersymmetric result.}
			\label{fig:specCompare}
		\end{minipage}
		\hfill
		\begin{minipage}[t]{\figWidth}
			\rule{0mm}{\picHeight}%
			\includegraphics[width=\linewidth]{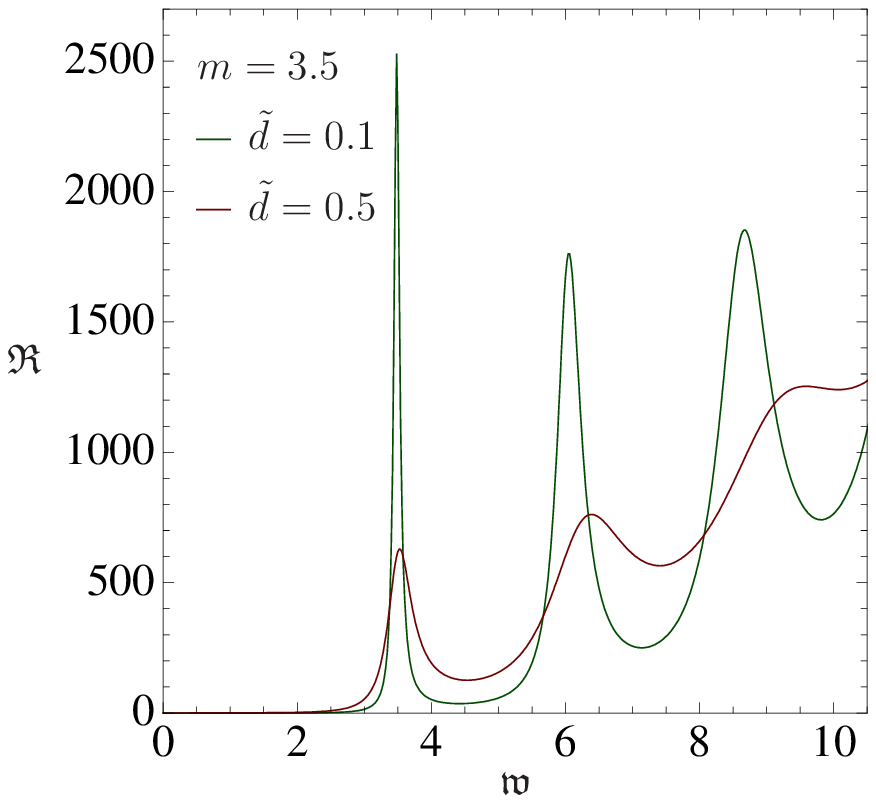}
			\caption{At increasing baryon density the resonances are broadened and
				slightly shifted.}
			\label{fig:specDensities}
		\end{minipage}
		\hfill
		\begin{minipage}[t]{\figWidth}
			\rule{0mm}{\picHeight}%
			\includegraphics[width=\linewidth]{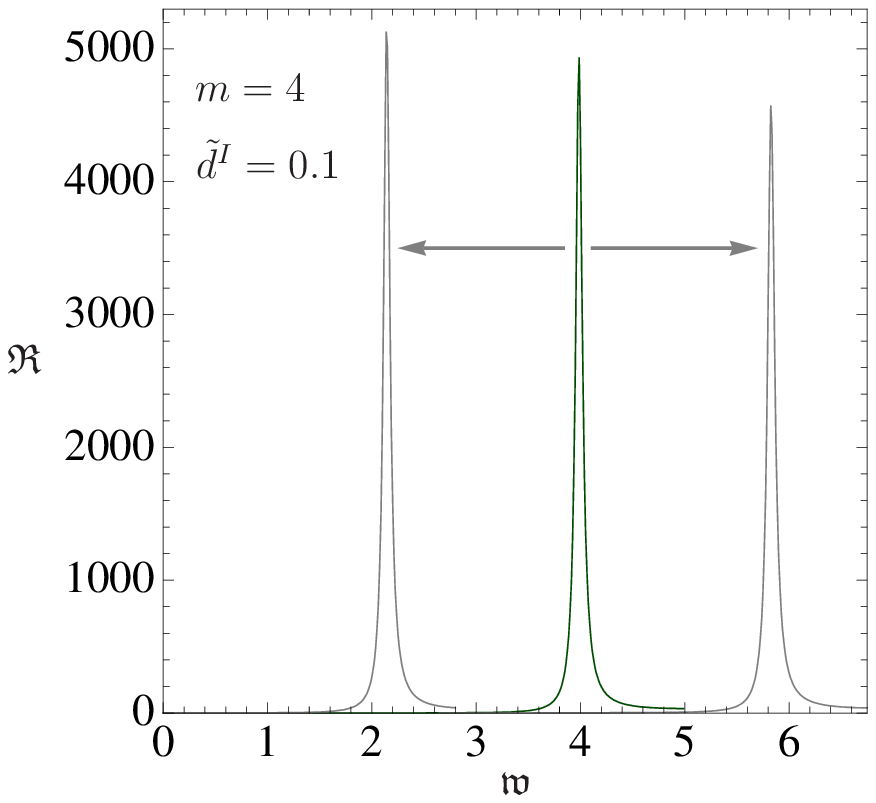}
			\caption{At finite isospin density the peaks split up to form an isospin
				triplet.}
			\label{fig:specSplit}
		\end{minipage}
	\end{figure}

	The resulting vector meson spectra depending on the normalized frequency
	$\mathfrak w=\omega/(2\pi T)$ for $N_f=2$ are shown in
	figures~\ref{fig:specCompare},\ref{fig:specDensities}
	and~\ref{fig:specSplit}.
	In figure~\ref{fig:specCompare} we compare the obtained spectra for a
	certain choice of quark mass/temperature ratio and particle density to the
	analytically known line spectrum for the case of zero temperature and
	vanishing particle density, where supersymmetry is restored. At low
	temperatures we observe resonance peaks in the derived spectra in accordance
	with the expectations from supersymmetric setups. Moreover we also see the
	phenomenologically expected behavior of finite resonance peak widths,
	indicating finite lifetimes of the bound states at finite temperature,
	referred to as \emph{meson melting}. Figure~\ref{fig:specDensities} shows
	the influence of variations in the particle density. Increasing the density
	destabilizes the bound states and slightly shifts the positions of the
	resonance peaks to higher energies. Such in-medium effects have also been
	observed in effective field theory models \cite{Rapp:1999us}. We do
	\emph{not} observe Brown-Rho scaling, which predicts decreasing meson masses
	\cite{Brown:1991kk}. Finally, figure~\ref{fig:specSplit} shows the splitting
	of the vector meson spectrum at finite isospin density. The isospin $1/2$
	quarks combine to the different possible combinations constituting an
	isospin $1$ meson triplet with $z$ components $-1$, $0$ and $+1$,
	respectively.
	
\section{Quarkonium diffusion}
	
	In this section we summarize the results of \cite{Dusling:2008tg} obtained
	in collaboration with K. Dusling, D. Teaney, and C. Young. We make use of
	the effective Lagrangian from \cite{Luke:1992tm} for the heavy meson field
	$\phi$ with mass $M$. Transcription to $\N=4$ SYM theory in lowest order in
	$T/M$ yields
	\begin{equation}
		\label{eq:effectiveLagr}
		\L = -i\, \phi^\dagger\del\phi + \frac{c_T}{N^2}\,\phi^\dagger \T^{00}\phi + \frac{c_F}{N^2}\,\phi^\dagger\mathcal O_{F^2} \phi.
	\end{equation}
	We refer to the last two terms as the interaction Lagrangian $\L_\text{int}$,
	where $\T^{00}=\frac38\,\pi^2N^2T^4$ is the $00$-component of the energy momentum
	tensor and $\mathcal O_{F^2}=\mathop\mathrm{tr}F^2$ is the operator which
	mediates the coupling of the mesons to the gauge bosons. The coefficients
	$c_T$ and $c_F$ determine the mass shift of the mesons in medium by
	$\delta M = -\left<\L_\text{int}\right>=-(c_E\langle\T^{00}\rangle+c_F\langle\mathcal O_{F^2}\rangle)/N^2$.
	
	In a Langevin process the diffusion of the mesons in the medium can be
	described by the momentum broadening $\kappa=2\pi MT\eta$, which determines
	the change of momentum $\vec p$ by damping and a random
	force $\vec\xi$ (random kicks) by
	\begin{equation}
		\frac{\dd p_i}{\dd t}=\xi_i(t)-\eta p_i,\quad\text{with}\quad\left<\xi_i(t)\,\xi_j(t')\right>=\kappa\,\delta_{ij}\delta(t-t').
	\end{equation}
	On time scales short compared to equilibration times but long compared
	to medium correlations we have
	\begin{equation}
	\label{eq:kappaFormula}
		\kappa \propto \int\dd t\int\dd t' \left<\xi_i(t)\xi_j(t')\right>=\int\dd t \left<\frac{\dd p_i}{\dd t}\frac{\dd p_j}{\dd t}\right>=\int\dd t\left<\mathcal F_i \mathcal F_j\right>.
	\end{equation}
	We derive the force $\mathcal F$ from the interaction Lagrangian $\mathcal
	L_\text{int}$, given by the last two terms in \eqref{eq:effectiveLagr},
	$\mathcal F=-\int\dd^3x\,\phi^\dagger\nabla\mathcal L_\text{int}\phi$.
	Plugging this into \eqref{eq:kappaFormula} leads to
	\begin{equation}
		\kappa = -\lim_{\omega\to\infty}\int\frac{\dd^3 q}{(2\pi)^3}\frac{2Tq^2}{3\omega N^4}
		\left(c_T^2\,\Im G^R_{\T} + c^2_F\,\Im G^R_{F^2} \right).
	\end{equation}
	The two-point correlation functions $G^R$ of the energy momentum tensor and
	the field strength tensor can be obtained perturbatively at weak coupling,
	and via \ads/CFT at strong coupling in a similar way as above. The same
	applies for the mass shift $\delta M$. The quotient $\kappa/(\delta M)^2$
	does not depend on the coefficients $c_E$ and $c_F$. We can compare this
	quantity in the limit of weak coupling with the result at asymptotically
	strong coupling \cite{Dusling:2008tg},
	\begin{equation}
		\left[ \frac{\kappa}{(\delta M)^2}\right]_{\lambda\to 0}=\frac{\pi T}{N^2}\,37.0\:,
		\qquad\qquad
		\left[ \frac{\kappa}{(\delta M)^2}\right]_{\lambda\to\infty}=\frac{\pi T}{N^2}\,8.3\:.
	\end{equation}
	We thus see that the momentum broadening is reduced in the limit of strong
	coupling, leading to increased diffusion and relaxation time.
	
\section{Conclusion}
	
	Using the example of the D3/D7 system, we showed that holographic models are
	capable of describing spectra of meson-like bound states of supersymmetric
	fundamental degrees of freedom. These spectra exhibit many qualitative
	features expected in finite temperature, finite density strongly coupled
	QCD. Among these are increased decay rates at high temperature, in-medium
	effects of increased decay rates and mass shifts at finite particle density,
	as well as the triplet splitting of vector meson spectra at finite isospin
	density.
	
	In an effective Lagrangian model we used holographic techniques to estimate
	strong coupling effects on meson diffusion in a model of the strongly
	coupled quark-gluon plasma. The results suggest that the drag forces on
	mesons in the medium are reduced at strong coupling, such that momentum
	broadening in the medium is reduced and relaxation times increase. In this
	sense the observed suppression of the quarkonium drag coefficient is analog
	to the suppression of the shear viscosity coefficient at strong coupling.

\section*{Acknowledgments}

	We are grateful to D. Teaney, K. Dusling and C. Young for fruitful
	collaboration. We thank the organizers of the 8th Conference on \emph{Quark
	Confinement and the Hadron Spectrum} for an enriching congress. Part of this
	work was funded by the \emph{Cluster of Excellence for Fundamental
	Physics\,---\,Origin and Structure of the Universe}.


\providecommand{\href}[2]{#2}\begingroup\raggedright\endgroup

\end{document}